# Title: Valley-Hall photonic topological insulators with dual-band kink states

*Qiaolu Chen, Li Zhang, Mengjia He, Zuojia Wang, Xiao Lin, Fei Gao, Yihao Yang [*], Baile Zhang[*], and Hongsheng Chen[*]*

Q. Chen, L. Zhang, M. He, Prof. F. Gao, Prof. H. Chen
State Key Laboratory of Modern Optical Instrumentation, College of Information Science and Electronic Engineering, Zhejiang University, Hangzhou 310027, China.
hansomchen@zju.edu.cn (H.C.)

Q. Chen, L. Zhang, M. He, Prof. F. Gao, Prof. H. Chen
Key Lab. of Advanced Micro/Nano Electronic Devices & Smart Systems of Zhejiang, The Electromagnetics Academy at Zhejiang University, Zhejiang University, Hangzhou 310027, China.

Prof. Z. Wang
School of Information Science and Engineering, Shandong University, Jinan 250100, China

Dr. X. Lin, Dr. Y. Yang, Prof. B. Zhang
Division of Physics and Applied Physics, School of Physical and Mathematical Sciences, Nanyang Technological University, 21 Nanyang Link, Singapore 637371, Singapore.
yang.yihao@ntu.edu.sg (Y.Y.)
blzhang@ntu.edu.sg (B.Z.)

Dr. Y. Yang, Prof. B. Zhang
Centre for Disruptive Photonic Technologies, The Photonics Institute, Nanyang Technological University, 50 Nanyang Avenue, Singapore 639798, Singapore.

**Keywords:** valley-Hall, dual-band, kink states, photonic topological insulators

**Abstract**

Extensive researches have revealed that valley, a binary degree of freedom (DOF), can be an excellent candidate of information carrier. Recently, valley DOF has been introduced into photonic systems, and several valley-Hall photonic topological insulators (PTIs) have been experimentally demonstrated. However, in the previous valley-Hall PTIs, topological kink states only work at a single frequency band, which limits potential applications in multiband waveguides, filters, communications, and so on. To overcome this challenge, here we experimentally demonstrate a valley-Hall PTI, where the topological kink states exist at two separated frequency bands, in a microwave substrate-integrated circuitry. Both the simulated and experimental results demonstrate the dual-band valley-Hall topological kink states are robust against the sharp bends of the internal domain wall with negligible inter-valley scattering. Our work may pave the way for multi-channel substrate-integrated photonic devices with high efficiency and high capacity for information communications and processing.

**Introduction**

Over the past few years, 'valleytronics'[1-5] has emerged as an area where valley, a binary degree of freedom, has a potential to be an excellent candidate of information carrier. The concept of valley has also been introduced into different kinds of physical systems, such as photonics,[6-13] acoustics,[14-17] and elastics.[18-20] In the photonic community, researchers have found that domain walls between two types of inversion-breaking photonic crystals, i.e., valley-Hall photonic topological insulators (PTIs),[9,11] could support topologically non-trivial valley-polarized kink states. Such valley-Hall PTIs have been experimentally demonstrated at different frequencies and found potential applications in topologically protected refraction,[8] high-efficiency waveguides,[6,11,12] topological waveguide splitters,[12] to name but a few. Whereas, in the previous valley-Hall PTIs, topological kink states only exist at a single frequency band, limiting multi-functional applications, such as multiband communications,[21] multiband filters,[22] multiband subwavelength resonators,[23] and so forth.

To overcome the above challenge, in this work, we propose and experimentally realize a valley-Hall PTI with dual-band kink states. The valley-Hall PTI is designed with an inversion-breaking graphene-like structure in a substrate-integrated microwave circuitry. At a domain wall between two valley-Hall PTIs with opposite effective masses, the topological kink states exist at two separated frequency bands. Such dual-band topological kink states are robust against sharp corners, as demonstrated in both simulations and experiments. Our work may lead to a number of potential multiband photonic devices and applications with high capacity for information storage and processing.

**Results**

The designed valley-Hall PTI is in a hexagonal lattice with lattice constant of $a = 16.45$ mm, as illustrated in Figure 1(a). Each unit cell (insets in Figure 1(a)) includes two cylindrical metallic scatters with radius of $r = 1.9$ mm and thickness of $h = 3.2$ mm, and a metallic mesh pattern on their top. For a unit cell, the top metallic mesh pattern includes two disks with different radii, $r_A$ and $r_B$, respectively, connected by a metal wire with width of $w = 1.65$ mm. In our work, the symmetry and topological phase transition of photonic crystals are controlled by varying the radii of metallic disks. The whole structure is arranged between two parallel metallic plates, loaded with a dielectric material with relative permittivity of $\varepsilon = 3.2$. The cylindrical metallic scatters touch the bottom metallic plate, and there is a gap with distance of $g_0 = 0.8$ mm between metallic mesh patterns and top metallic plate. In our implementation, the experimental sample is composed of two printed circuit boards (PCBs) as shown in Figure 1(b). The metallic mesh patterns and via holes are printed on the bottom PCB. The top copper PCB provides a gap and is attached on the bottom PCB. It is clear that our structure has advantages of easy access, excellent electromagnetic shielding, and compatibility with conventional substrate-integrated microwave circuitry.

When $r_A = r_B$, the photonic crystal possesses inversion symmetry and time-reversal symmetry.[13,14] As shown in Figure 2(a), there are two pairs of Dirac-like degenerate points in the first Brillouin zone at the frequency of 5.34 GHz and 14.11 GHz, respectively. The numerical dispersions are obtained by employing the eigenvalue module of the commercial software COMSOL Multiphysics.

When slightly breaking the inversion symmetry by changing the radii $r_A$ and $r_B$ to make them non-equivalent, while keeping $r_A + r_B$ a constant, the symmetry of hexagonal lattice is reduced from $C_{6v}$ to $C_3$. This perturbation renders the Dirac points at K (or K') valley lifted and hence two bandgaps open at the previous Dirac frequencies. For example, when $\Delta r = r_A - r_B = 1.2$ mm, two bandgaps appear, whose operational frequency ranges vary from 4.94 GHz to 5.74 GHz for the first bandgap, and from 13.35 GHz to 14.12 GHz for the second bandgap, respectively, as shown in Figure 2(b). Note that though the gap between metallic mesh patterns and top metallic plate breaks the $z$-inversion symmetry, the designed PTI differs from bi-anisotropy induced spin-Hall PTIs that involve both transverse-electric (TE) and transverse-magnetic (TM) waves.[24-26] Because the height of parallel waveguide is only 3.2 mm, which is less than half of the operational wavelength, only TE modes survive.

We also plot the time-averaged Poynting vectors of the four eigenstates at K valley on the $z$=3.27 mm plane when the inversion symmetry is broken, as shown in Figure 2(d). One can see that the energy flux of the four eigenstates at K valley are either left circularly polarized (LCP) or right circularly polarized (RCP). Applying the time-reversal operation, we could also obtain similar energy flux of the four eigenstates at K' valley.

According to $\mathbf{k} \cdot \mathbf{p}$ method, the effective Hamiltonian around K/K' valley can be expressed as $H_{K/K'}(\delta k) = \pm\left(v_D \delta k_x \sigma_x + v_D \delta k_y \sigma_y\right) \pm m v_D^2 \sigma_z$, where $\delta \mathbf{k} = \mathbf{k} - \mathbf{k}_{K/K'}$ is the displacement of wave vector $\mathbf{k}$ to K/K' valley in the reciprocal space, $v_D$ is the group velocity, $m$ is the effective mass term, and $\sigma_i = (i=x, y, z)$ are elements in the Pauli matrices.[6,8,9,14] Solving the effective Hamiltonian, we can obtain the Berry curvature at K/K' valley, $\Omega_{K/K'}(\delta k) = \pm\dfrac{m v_D}{2(\delta k^2 + m^2 v_D^2)^{3/2}}$.[6,14] In order to consolidate the above theoretical analysis, we calculate the Berry curvatures of the four bands near K valley by performing first-principle simulations of the practical structure in the COMSOL Multiphysics, as shown in Figure 2(e). Note that the Berry curvatures near K' valley, which are not shown here, are opposite to those near K valley. Integrating the Berry curvature around K/K' valley, we obtain the valley-Chern number at K/K' valley analytically,[6,8,9,14] $C_{K/K'} = \pm\dfrac{1}{2}\text{sgn}(m)$, which is consistent with our first-principle simulation. The mass term $m$ is characterized by the difference between the frequency of RCP and that of LCP, and is adjusted by altering $\Delta r$ in our case. Sweeping $\Delta r$, we obtain the phase diagram as shown in Figure 2(c), which exhibits a topological phase transition when $\Delta r$ crosses zero.[6,14]

Then we construct a 'kink'-type domain wall between two valley-Hall PTIs with opposite mass terms (or opposite $\Delta r$). Across the domain wall, the valley-Chern number varies from $\pm 1/2$ to $\mp 1/2$ at K/K' valley, resulting in $\Delta C = \pm 1$. Therefore, according to the bulk-edge correspondence, two topological kink states appear at the domain wall around K/K' valley.[6,8,11,12,14] Note that the above analysis is applicable for both pairs of Dirac points at different frequencies, therefore, topological kink states exist at both bandgaps.

To verify the above statement, we perform simulations for two 'kink'-type domain walls consisting of two valley-Hall PTIs with opposite valley-Chern numbers (or opposite $\Delta r$), as shown in Figure 3(a) and 3(b). The kink states at two domain walls exhibit different spatial symmetries: even-symmetry (Figure 3(a)) or odd-symmetry (Figure 3(b)). The dispersions of dual-band topological kink states are shown in Figure 3(c) and 3(d), where the gray regions and lines represent the projections of bulk bands and the dispersions of kink states, respectively. Considering a certain bandgap, there is only a single kink state for each valley. Moreover, the propagation directions of kink states at different valleys are precisely opposite, exhibiting "valley-locked" chirality.[14] Such a phenomenon exists at two separated frequency bands, resulting in dual-band topological kink states. As the dual-band topological kink states are locked to K/K' valley, the inter-valley scattering is strongly suppressed in the presence of disorders, for example, sharp corners. Such a property makes the designed valley-Hall PTI to be an excellent candidate for wave-guiding.

To verify the robustness of the proposed topological kink states at two separated frequency bands, we design and compare two types of domain walls: straight (Figure 4(a)) and z-shape domain walls (Figure 4(d)). As shown in Figure 4(b) and 4(c), the $E_z$ field distributions depict that energy is strongly confined at the straight domain wall when the operational frequency is within the first or second bandgap. We also simulate the $E_z$ field distributions around the z-shape domain wall featuring two sharp corners (Figure 4(e) and 4(f)). One can see that the topological kink states smoothly go through both of sharply twisted corners (120° turn) with negligible reflections at two separated frequency bands.

In order to further experimentally prove the high-transmission property of the designed photonic topological waveguides, we measure the transmissions of kink states around two topological bandgaps for three cases (Figure 4(g) and 4(h)): without domain wall (black lines), with straight domain wall (blue lines), and with z-shape domain wall (red lines). As expected, for the case without domain wall, there are two obvious drops from 5.0 GHz to 6.1 GHz, and from 12.9 GHz to 14.5 GHz, revealing the presence of dual bandgaps. Besides, the transmissions of experimental sample with straight domain wall are approximately 30 dB higher than those without domain wall at both bandgaps, which indicates the existence of dual-band topological kink states. Meanwhile, the transmissions along the z-shape domain wall are almost identical to those along the straight domain wall, manifesting the dual-band topological kink states are robust against sharp corners. It is clear that experimental observations are consistent with the simulated results, confirming our theoretical prediction that the dual-band valley-Hall topological kink states are robust against the sharp bends, and proving the topological protection of the kink states.

**Conclusions**

In summary, we experimentally demonstrate a valley-Hall PTI with topological kink states at two separate frequency bands in a microwave substrate-integrated circuitry. Both the simulated and experimental results verify that the dual-band topological kink states are robust against the sharp bends with negligible inter-valley scattering. Our valley-Hall PTI with dual-band kink states has advantages of self-consistent electrical shielding, easy access, and comparability with conventional substrate-integrated waveguide circuitry, which may find applications in multi-channel wave-guiding and communications. Moreover, though the design principle of valley-Hall PTIs with dual-band kink states is verified at microwave frequency, its universality could be applied in terahertz and even optical regimes.

**Acknowledgments**
Q.C. and Y.Y. contributed equally to this work. This work was sponsored by the National Natural Science Foundation of China under Grants No. 61625502, No. 61574127, and No. 61801426, the Top-Notch Young Talents Program of China, the Fundamental Research Funds for the Central Universities, and the Innovation Joint Research Center for Cyber-Physical-Society System.

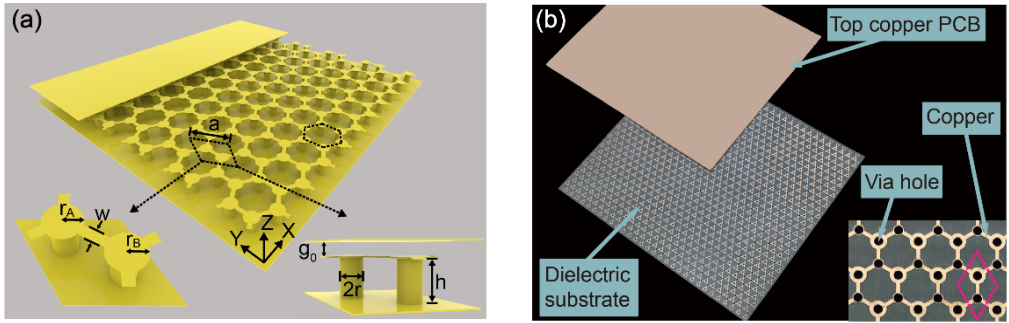

**Figure 1. Schematic view of the valley-Hall PTI with dual-band kink states. (a)** Structure of the valley-Hall PTI. Dashed black rhombic and hexagonal lines label the unit cell. Insets: details of rhombic unit cell. $a$ is the lattice constant; $r$ and $h$ are the radius and thickness of cylindrical metallic scatters, respectively; $g_0$ is the distance of the gap; $r_A$ and $r_B$ are the radii of two adjacent metallic disks, respectively; $w$ is the width of wires connecting adjacent metallic disks. Here, $a$ = 16.45 mm, $r$ = 1.9 mm, $w$ = 1.65 mm, $g_0$ = 0.8 mm, $h$ = 3.2 mm, and $r_A + r_B$ = 5.75 mm. The thicknesses of the metal layers are 0.035 mm. **(b)** Perspective-view photograph of the experimental sample fabricated with PCB technology. Inset: enlarged view of experimental structure. Red dashed rhombic line depicts a unit cell.

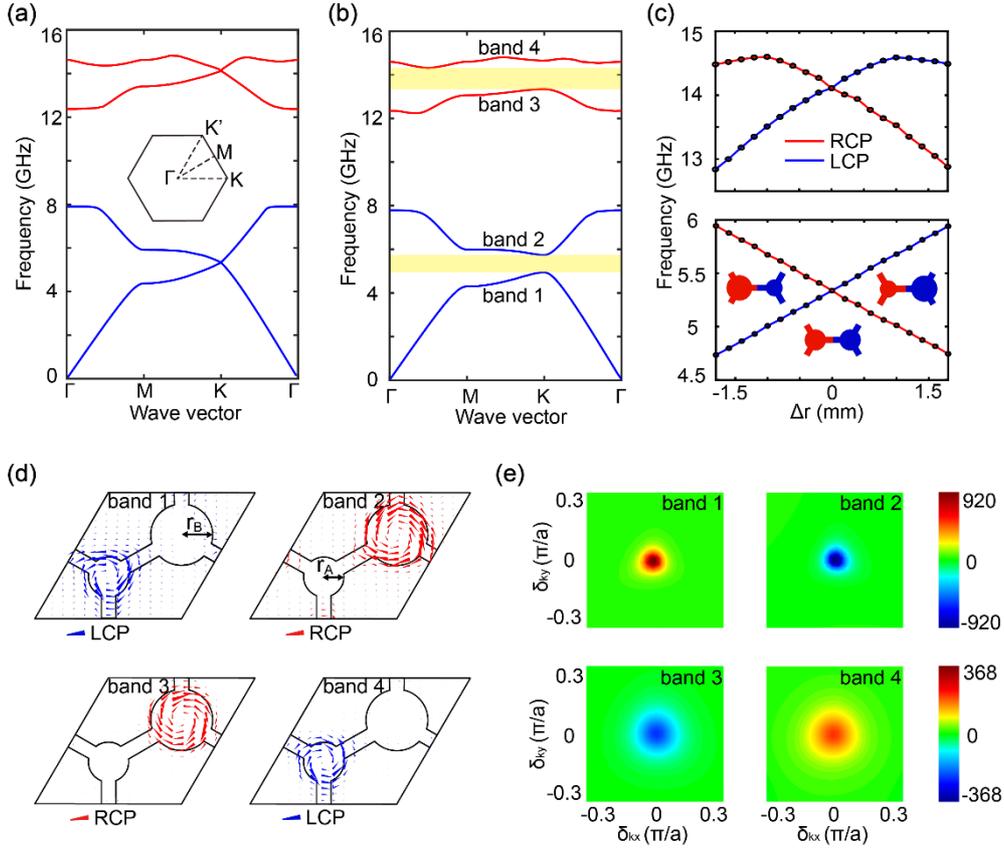

**Figure 2. Band structures, mode analysis, and Berry curvatures of the valley-Hall PTIs with dual-band kink states.** **(a)** Band structures when $\Delta r = 0$. Inset: the first Brillouin zone of hexagonal lattice. **(b)** Band structures when $\Delta r = 1.2$ mm. Yellow areas represent the bandgaps. **(c)** Phase diagram of the eigenmodes at K valley when the parameter $\Delta r$ varies from -1.8 mm to 1.8 mm while $r_A + r_B$ keeps a constant. Insets: unit cells when $\Delta r = -1.2$ mm (left inset), $\Delta r = 0$ (middle inset), and $\Delta r = 1.2$ mm (right inset), respectively. Red lines: RCP modes. Blue lines: LCP modes. **(d)** Simulated time-averaged Poynting vectors at K valley on the $z$=3.27 mm plane of the four bands after inversion-symmetry breaking. Blue (red) arrows indicate LCP (RCP) modes. **(e)** Local Berry curvatures (cm$^2$) of the four bands near K valley in the reciprocal space.

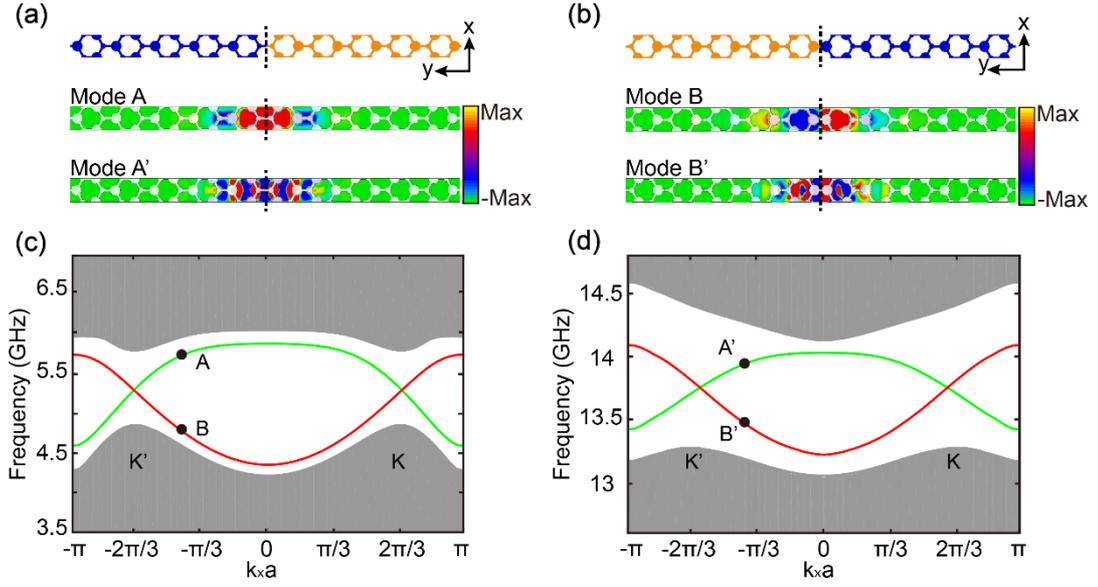

**Figure 3. Dispersion relations of two 'kink'-type domain walls consisting of opposite symmetry-breaking geometries.** **(a), (b)** Two 'kink'-type domain walls formed by two valley-Hall PTIs with opposite symmetry-breaking geometries, $\Delta r = 1.2$ mm on the left (right) and $\Delta r = -1.2$ mm on the right (left). The field patterns represent the $E_z$ field distributions of mode A and A' (even-symmetry), mode B and B' (odd-symmetry) labeled in **(c)** and **(d)**, respectively. **(c), (d)** Band structures of domain walls at two separated frequency bands. Green lines: valley-Hall topological kink states of the domain wall in **(a)**. Red lines: valley-Hall topological kink states of the domain wall in **(b)**. Grey areas: projected bulk states.

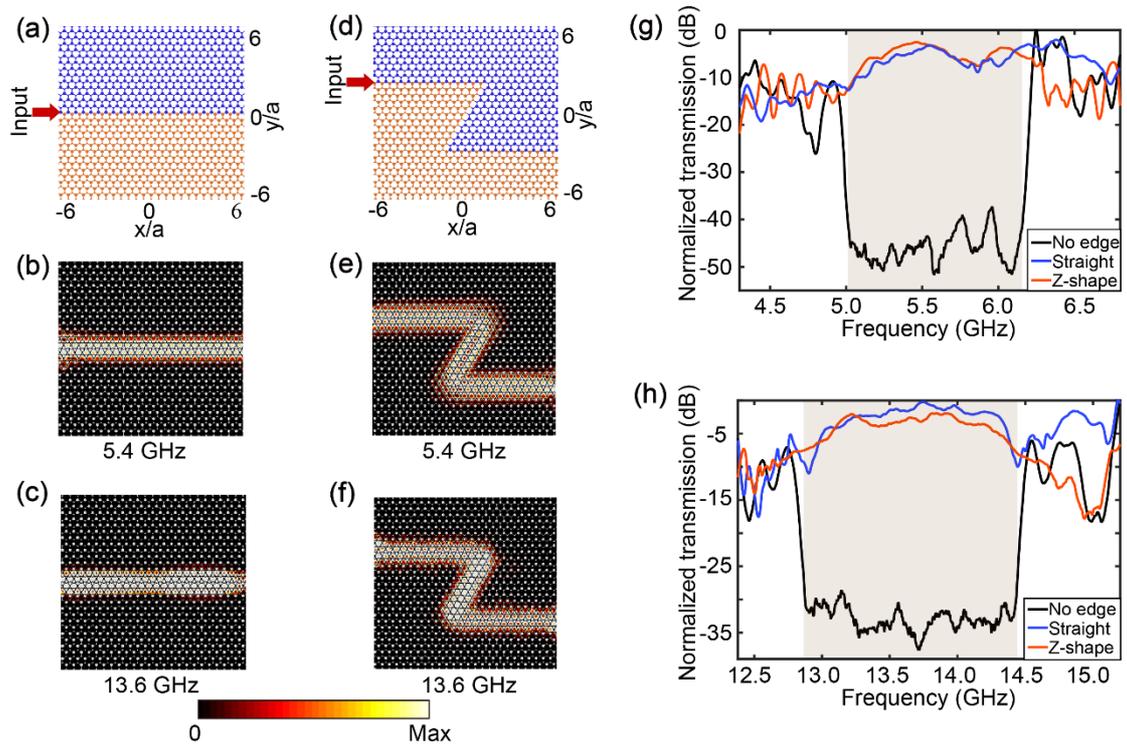

**Figure 4. Experimental demonstration of robust valley-Hall topological kink states at two separated frequency bands. (a), (d)** Schemes of straight and z-shape domain walls. The excitations are on the left. The orange (blue) hexagonal lattice represents the structures with positive (negative) $\Delta r$. **(b), (e)** Simulated electric field intensity distributions at 5.4 GHz corresponding to **(a)** and **(d)**, respectively. **(c), (f)** Simulated electric field intensity distributions at 13.6 GHz corresponding to **(a)** and **(d)**, respectively. The color depicts the amplitudes of $E_z$. **(g), (h)** Normalized measured transmissions of kink states around two topological bandgaps for three cases: without domain wall (black), with straight domain wall (blue), and with z-shape domain wall (red). Grey areas: bandgaps. The experimental results are normalized by the maximum values of the corresponding bulk transmissions.